\documentclass[letter]{emulateapj}
\usepackage{apjfonts,psfig,epsfig}

\lefthead{Smith, G.\ P., et al.}
\righthead{MACS\,J1149.5+2223}

\begin{document}
\def\kpc{\mathrel{\rm kpc}}
\def\Msol{\mathrel{\rm M_{\odot}}}
\def\fsub{\mathrel{f_{\rm sub}}}
\def\Mtot{\mathrel{M_{\rm tot}}}
\def\ls{\mathrel{\hbox{\rlap{\hbox{\lower4pt\hbox{$\sim$}}}\hbox{$<$}}}}
\def\gs{\mathrel{\hbox{\rlap{\hbox{\lower4pt\hbox{$\sim$}}}\hbox{$>$}}}}
\def\Msolpyr{\mathrel{\rm M_{\odot}\,yr^{-1}}}
\def\mas{\mathrel{\rm mas}}
\def\pc{\mathrel{\rm pc}}
\def\Ho{\mathrel{H_{\rm 0}}}
\def\oM{\mathrel{\Omega_{\rm M}}}
\def\oL{\mathrel{\Omega_{\rm \Lambda}}}
\def\kms{\mathrel{\rm km\,s^{-1}}}
\def\Mpc{\mathrel{\rm Mpc}}
\def\ksec{\mathrel{{\rm ksec}}}
\def\mag{\mathrel{\rm mag}}
\def\Gyr{\mathrel{\rm Gyr}}
\def\dls{\mathrel{D_{\rm LS}}}
\def\dos{\mathrel{D_{\rm OS}}}
\def\dol{\mathrel{D_{\rm OL}}}
\def\zl{\mathrel{z_{\rm L}}}
\def\zs{\mathrel{z_{\rm S}}}
\def\pix{\mathrel{\rm pixel}}
\def\AA{\mathrel{\hbox{\rlap{\hbox{\raise6.5pt\hbox{{\hspace{0.5mm}\tiny$\circ$}}}}\hbox{\small A}}}}

\title{\emph{Hubble Space Telescope Observations of a Spectacular New
    Strong-lensing Galaxy Cluster -- MACS\,J1149.5+2223 at z=0.544}}

\slugcomment{Received 2009 June 29; Accepted 2009 November 4}

\author{
Graham P.\ Smith,$\!$\altaffilmark{1,2}
Harald Ebeling,$\!$\altaffilmark{3}
Marceau Limousin,$\!$\altaffilmark{4,5}
Jean-Paul\ Kneib,$\!$\altaffilmark{4}
A.\ M.\ Swinbank,$\!$\altaffilmark{6}
Cheng-Jiun Ma,$\!$\altaffilmark{3}
Mathilde Jauzac,$\!$\altaffilmark{4}
Johan Richard,$\!$\altaffilmark{6}
Eric Jullo,$\!$\altaffilmark{7}
David J.\ Sand,$\!$\altaffilmark{8,9}
Alastair C.\ Edge,$\!$\altaffilmark{6}
Ian Smail$\!$\altaffilmark{6}
}

\altaffiltext{1}{School of Physics and Astronomy, University of
  Birmingham, Edgbaston, Birmingham, B15 2TT, UK.  Email: 
  gps@star.sr.bham.ac.uk}
\altaffiltext{2}{California Institute of Technology, Mail Code 105-24, 
  Pasadena, CA 91125, USA} 
\altaffiltext{3}{Institute for Astronomy, University of Hawaii, 
  2680 Woodlawn Drive, Honolulu, HI 96822, USA}
\altaffiltext{4}{Laboratoire d'Astrophysique de Marseille,
  CNRS-Universit\'e Aix-Marseille, 38 rue F.\ Joliot-Curie, 13388 
  Marseille Cedex 13, France}
\altaffiltext{5}{Dark Cosmology Centre, Niels Bohr Institute,
  University of Copenhagen, Juliane Maries Vej 30, 2100 Copenhagen,
  Denmark} 
\altaffiltext{6}{Institute for Computational Cosmology, Durham
  University, South Road, Durham, DH1 3LE, UK} 
\altaffiltext{7}{Jet Propulsion Laboratory, California Institute of
  Technology, MS 169-506, Pasadena, CA 91125} 
\altaffiltext{8}{Harvard-Smithsonian Center for Astrophysics, 60
  Garden Street, Cambridge, MA 02138, USA}
\altaffiltext{9}{Harvard Center for Astrophysics and Las Cumbres
  Observatory Global Telescope Network Fellow}

\begin{abstract}
  We present Advanced Camera for Surveys observations of
  MACS\,J1149.5$+$2223, an X-ray luminous galaxy cluster at
  $z{=}0.544$ discovered by the Massive Cluster Survey.  The data
  reveal at least seven multiply-imaged galaxies, three of which we
  have confirmed spectroscopically.  One of these is a spectacular
  face-on spiral galaxy at $z=1.491$, the four images of which are
  gravitationally magnified by $8\ls\mu\ls23$.  We identify this as an
  $L^\star$ ($M_B\simeq-20.7$), disk-dominated ($B/T\ls0.5$) galaxy,
  forming stars at $\sim6\Msolpyr$.  We use a robust sample of
  multiply-imaged galaxies to constrain a parameterized model of the
  cluster mass distribution.  In addition to the main cluster dark
  matter halo and the bright cluster galaxies, our best model includes
  three galaxy-group-sized halos.  The relative probability of this
  model is ${\rm P}(N_{\rm halo}=4)/{\rm P}(N_{\rm halo}<4)\ge10^{12}$
  where $N_{\rm halo}$ is the number of cluster/group-scale halos.  In
  terms of sheer number of merging cluster/group-scale components,
  this is the most complex strong-lensing cluster core studied to
  date.  The total cluster mass and fraction of that mass associated
  with substructures within $R\le500\kpc$, are measured to be
  $\Mtot=(6.7\pm0.4)\times10^{14}\Msol$ and $\fsub=0.25\pm0.12$
  respectively.  Our model also rules out recent claims of a flat
  density profile at $\gs7\sigma$ confidence, thus highlighting the
  critical importance of spectroscopic redshifts of multiply-imaged
  galaxies when modeling strong lensing clusters.  Overall our results
  attest to the efficiency of X-ray selection in finding the most
  powerful cluster lenses, including complicated merging systems.
\end{abstract}


\section{ Introduction }\label{sec:intro}

Strong gravitational lensing by galaxy clusters is a well-established
probe of the extragalactic Universe, offering a magnified view of high
redshift galaxies that would otherwise be beyond the reach of present
day telescopes \citep[e.g.][]{Franx97, Ellis01, Kneib04b, Smail07,
  Swinbank07, Bradley08}.  The detailed cluster mass models required
to intepret such observations contain a wealth of information about
the mass and structure of cluster cores, against which theoretical
predictions can be tested \citep[e.g.][]{Smith05a,Comerford06,Sand08}.

The majority of spectroscopically confirmed strong lensing clusters
are at low redshift, i.e.\ $z{\ls}0.3$
\citep[e.g.][]{Kneib96,Broadhurst05acs,Smith05a,Limousin07a1689,Richard09}.
In contrast, only four spectroscopically confirmed strong lensing
clusters are known at $z{>}0.5$
\citep{Gladders02,Inada03,Borys04,Sharon05,Swinbank07,Ofek08}.  The
Massive Cluster Survey \citep[MACS;][]{Ebeling01} offers an
unprecedented opportunity to expand the available sample of strong
lensing clusters at $0.3\le z\ls0.7$.  We present new results from
this search: MACS\,J1149.5$+$2223 (hereafter MACS\,J1149; 11:49:34.3
$+$22:23:42.5 [J2000]) at $z=0.544$, one of a complete subsample of 12
MACS clusters at $z>0.5$ \citep{Ebeling07}.

In \S\ref{sec:obs} we describe the data; modeling and results are
presented in \S\ref{sec:results}, and summarized in
\S\ref{sec:summary}. We assume $\Ho=70\kms\Mpc^{-1}$, $\oM=0.3$ and
$\oL=0.7$; at $z=0.544$ $1''$ corresponds to $6.35\kpc$.  All
uncertainties and upper/lower limits are stated and/or plotted at
$95\%$ confidence.

\section{ Observations and Data Analysis }\label{sec:obs}

\begin{figure*}
\psfig{file=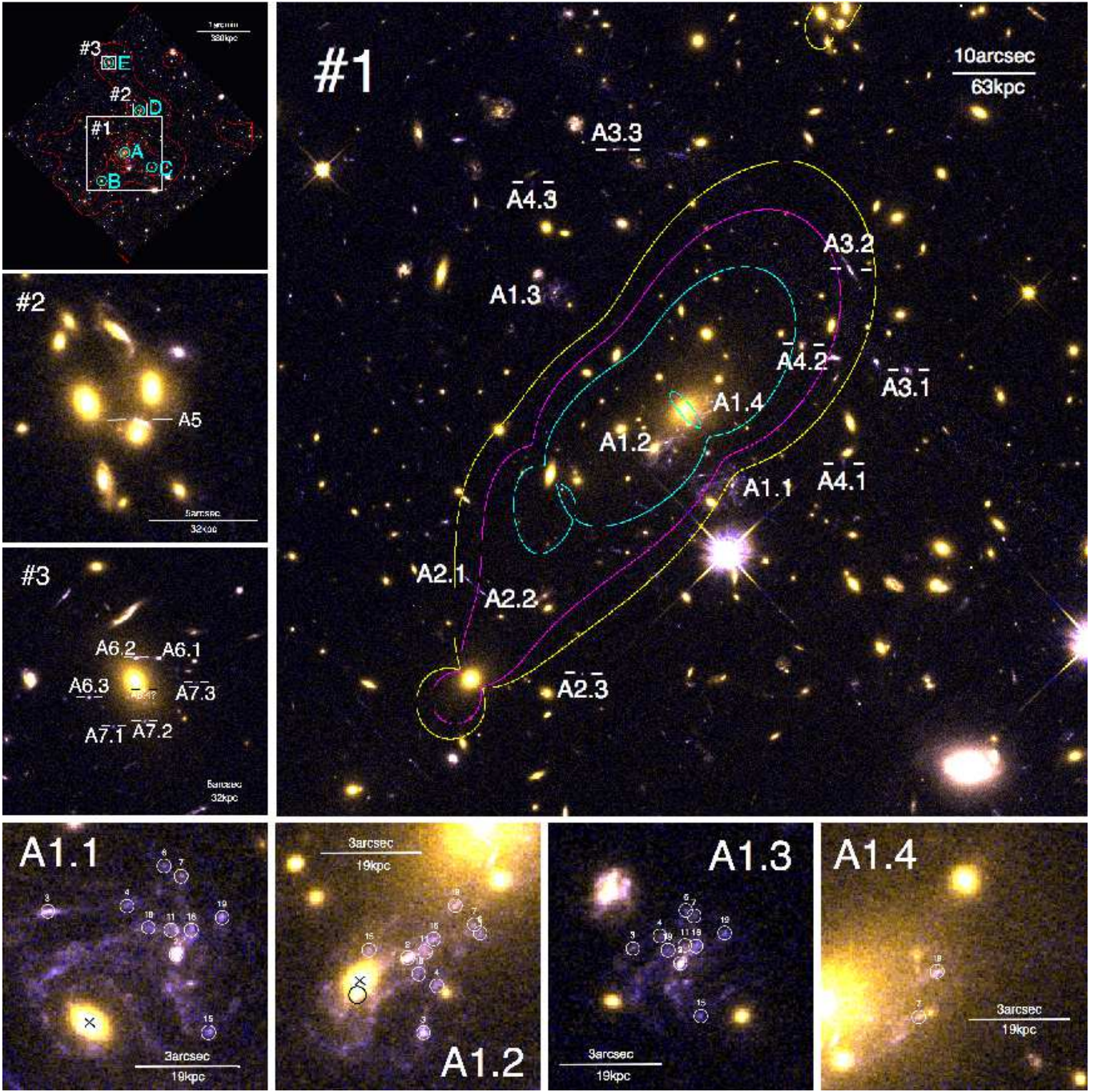,width=180mm,angle=0}
\caption{ {\sc Top left} -- $V_{555}/I_{814}$-band color picture
  showing the full ACS field of view.  Red contours show the
  luminosity density of cluster galaxies -- cluster halos were
  centered on the 5 luminous structures labeled A--E in the lens model
  (\S\ref{sec:model}).  The white boxes marked \#1, \#2 and \#3 show
  the regions displayed in more detail in the three numbered panels.
  {\sc Panel \#1} -- The central $\sim80''\times80''$ of the cluster
  showing the multiple image systems discussed in the text.  The cyan
  (outer), magenta and yellow curves show the $z=1.491$, $z=1.894$ and
  $z=2.497$ tangential critical curves respectively.  The inner cyan
  curve shows the radial critical curve for $z=1.491$.  {\sc Panel
    \#2} -- A faint triply-imaged galaxy next to a cluster galaxy
  within a group of galaxies $\sim50\arcsec$ North of the BCG.  {\sc
    Panel \#3} -- Two candidate triply-imaged systems adjacent to a
  bright cluster galaxy $\sim100\arcsec$ North of the BCG; A6.4 marks
  the location of a possible fourth image of A6.  {\sc Bottom row} --
  Zoom into the four images of A1; morphological features used to
  constrain the lens model are marked by numbered white circles.  The
  black crosses and circle in the A1.2/A1.1 panels are discussed in
  \S\S\ref{sec:model}~\&~\ref{sec:zitrin}.  N is up E is left in all
  panels.
\label{fig:arcs}
}
\end{figure*}

MACS\,J1149 was observed on 2004, April 22 with the Advanced Camera
for Surveys (ACS) on-board the \emph{Hubble Space Telescope
  (HST)}\footnote{Based in part on observations with the NASA/ESA
  \emph{Hubble Space Telescope} obtained at the Space Telescope
  Science Institute, which is operated by the Association of
  Universities for Research in Astronomy, Inc., under NASA contract
  NAS 5-26555.} for $4.5\ksec$ and $4.6\ksec$ through the F555W and
F814W filters respectively (GO:~9722, PI:~Ebeling).  These data were
reduced using standard {\sc multidrizzle} routines onto a
$0.03''/\!\pix$ grid.  The reduced data reveal a striking
multiply-imaged disk galaxy comprising three tangential images
(A1.1/2/3; Fig.~\ref{fig:arcs}), and an additional image (A1.4) likely
caused by part of the galaxy's disk lying adjacent to the radial
caustic in the source plane.  A blue image pair (A2.1/2) also lies
$\sim30''$ South East of the BCG, with its counter-image (A2.3)
$\sim15''$ to the South West.  A further five triply-imaged galaxies
are identified based on their distinctive colors and morphologies:
A3.1/2/3 and A4.1/2/3 both lie between the BCG and a dense group of
cluster ellipticals $\sim50''$ to the NNW; A5 is embedded in the halo
of a cluster elliptical in the same group to the NNW of the BCG;
A6.1/2/3 and A7.1/2/3 surround a bright elliptical galaxy $\sim100''$
North of the BCG.  Numerous other faint blue background galaxies can
be seen through the cluster core, however the lack of concordant
colors and morphologies preclude a reliable identification of them as
being multiply-imaged at this time.

\begin{figure}
  \psfig{file=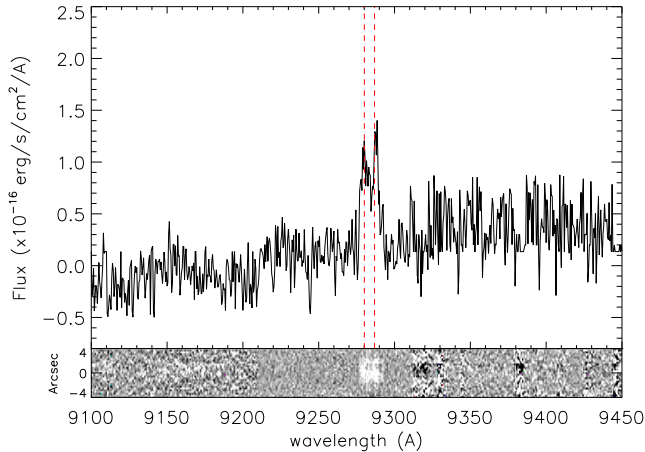,width=65mm,angle=90}
\caption{ Stacked one-dimensional and two-dimensional spectra of A1.1,
  A1.2 and A1.3.  The $1\arcsec$ wide slit on each target sampled both
  the central bulge and the disk of each image.
\label{fig:spect}
}
\end{figure}

MACS\,J1149 was observed with the Low Resolution Imaging Spectrograph
\citep[LRIS;][]{oke95} on the Keck-I 10-m telescope\footnote{The W.\
  M.\ Keck Observatory is operated as a scientific partnership among
  the California Institute of Technology, the University of
  California, and NASA.} on 2004, March 28 and 2005, March 6 employing
a single multi slit mask per run.  In 2004, we used the 400/3400
grism, D560 dichroic, and 400/8500 grating centered at $7200\AA$.  A
total integration of $10.8\ksec$ yielded the redshift of A2.1/2 as
$z{=}1.894$ via detection of Lyman-$\alpha$ in emission at $3519\AA$
(A2.3 was also confirmed at $z=1.894$ in 2005).  This redshift was
used to constrain a preliminary lens model from which the redshift of
A1 was predicted to be $z\simeq1.5\pm0.1$.  In March 2005 we centered
the 831/8200 grating at $8750\AA$ to search for redshifted [{\sc oii}]
from A1.  The resulting 1-hr spectrum resolved the [{\sc oii}] doublet
at $\lambda_{\rm obs}=9282.5\AA$ in all three of A1.1/2/3, placing
this galaxy at $z=1.4906\pm0.0002$ (Fig.~\ref{fig:spect}).  The 2004
observations also identified A3.1 and A3.2 at $z=2.497$ via detection
of the Lyman break, plus interstellar Silicon and Carbon absorption
lines.

\section{ Modeling and Results }\label{sec:results}

\subsection{ Luminosity Density Map }\label{sec:lum}

To gain an initial view of the structure of the cluster mass
distribution we selected galaxies with spectroscopic or photometric
redshifts between $0.5\le z\le 0.6$ from Ma et al.'s (in prep.; see
also \citeauthor{Ma08} \citeyear{Ma08}) catalog.  This catalog is
based on spectroscopic observations with DEIMOS on Keck II (yielding
217 cluster members), plus panoramic $B/V/R_c/I_c/z^\prime$-band
imaging with Suprime-CAM on the Subaru 8.2-m telescope and similar
$u^\ast$-band data from MegaPrime on the CFHT 3.6-m telescope.  The
resulting luminosity density map is adaptively smoothed to $3\sigma$
significance using {\sc asmooth} \citep{Ebeling06} and is presented in
Fig.~\ref{fig:arcs}.  It reveals five luminous structures (labeled
A--E), suggesting that the underlying distribution of dark matter may
be similarly complicated.

\subsection{ Gravitational Lens Model }\label{sec:model}

Our goal is to constrain the shape of the mass distribution in the
cluster core.  We therefore adopt stringent criteria for the inclusion
of multiple image systems as constraints on our lens model to guard
against detection of spurious features in the mass distribution.  To
be included, a galaxy or morphological feature within a galaxy must be
identified a minimum of three times, and the morphological and color
match between the multiple images of the galaxy/feature must be
unambiguous.  Ten morphological features of A1 satisfy these criteria,
of which eight are seen in all of A1.1, A1.2 and A1.3, and two have
also been identified in A1.4 (Fig.~\ref{fig:arcs}).  These 10 features
lie generally away from the portions of the disk that are affected by
the cluster ellipticals marked with a black cross in
Fig.~\ref{fig:arcs}.  Note that we interpret all of the morphological
features South-East of the central bulge (feature \#2) of A1.2 as
being part of the disk of that image, distorted by the neighboring
cluster elliptical.  The three images each of A2, A3, A4, A6, and A7
are also used as model constraints, with the unknown redshifts of the
latter three being free parameters in the lens model.  These systems
total $n_{\rm c}=61$ model constraints.

The mass distribution was initially parameterized as a superposition
of 21 cluster galaxies ($I_{814}<20.5$), plus 5 cluster-scale
components (hereafter referred to as halos) centered on the brightest
galaxy in each of the light concentrations marked in
Fig.~\ref{fig:arcs}.  All galaxies and halos were parameterized as
smoothly-truncated pseudo-isothermal elliptical mass distributions
(PIEMD) following \cite{Kneib96}.  The position, ellipticity, and
orientation of the galaxies were matched to those of their light, and
the velocity dispersions, core and cut-off radii were scaled with
their luminosity, adopting the best-fit parameters for an $L^\star$
galaxy obtained by \cite{Smith05a}.  We imposed a prior of $0.544<z<3$
on the redshifts of A4, A6, and A7, the upper limit coming from the
absence of an obvious Lyman break within/blueward of F555W for all
three galaxies.  In total the model has $n_{\rm p}=16$ free
parameters.

The model was fitted to the data using the Bayesian MCMC sampler
within {\sc lenstool v6.5}\footnote{{\sc lenstool} is available
  online: http://www.oamp.fr/cosmology/lenstool}
\citep{Kneib96,Jullo07}, using a positional uncertainty of
$0.4\arcsec$ in the image plane.  The image-plane $\chi^2$ of the
best-fitting model was $\chi^2_{\rm min}=55.1$, and the average root
mean square (rms) deviation of images predicted by this model from the
observed positions is $\langle\sigma_i\rangle=0.5\arcsec$.  Halo A has
a velocity dispersion of $\sigma\simeq1270\kms$, and halos B, D, and E
have $\sigma\simeq400-500\kms$.  However halo C has a $95\%$
confidence upper limit of $\sigma<343\kms$, implying that the galaxies
associated with halo C may not be embedded in an extended dark matter
halo.

To explore this further we re-fitted the model excluding halo C, again
obtaining $\chi^2_{\rm min}=55.1$ and
$\langle\sigma_i\rangle=0.5\arcsec$.  We then used the Bayesian
evidence, i.e.\ the probability of the model given the data and the
choice of the PIEMD paramaterization, to determine whether the
additional complexity of the 5 halo model is justified by the data.
The result is summarized in Table~\ref{tab:models} -- the probability
of the 4 halo model exceeds that of the 5 halo model by a factor of
$\sim20\times$.  We therefore conclude that halo C is not justified by
the data.  We also test whether even simpler models offer more
probable descriptions of the data -- results are listed in
Table~\ref{tab:models}.  In summary, models with $<4$ halos are less
probable than the four halo model by $\sim12-107$ orders of magnitude.
We therefore adopt the four halo model as our fiducial model, and list
its parameters in Table~\ref{tab:models}.  

\begin{table*}
\begin{center}
  \caption{Details of Gravitational Lens Models}
\begin{tabular}{lllccccccc}
\hline
\hline
\noalign{\smallskip}
   &          & ~~~ & ${\Delta}$RA   & ${\Delta}$Dec  & ${\epsilon}$           & ${\theta}$            & $\sigma$              & $r_{\rm core}$     & $r_{\rm cut}$     \\
~~ &    &                    & (arcsec)       & (arcsec)       &                        & (degrees)             & ($\kms$)              & ($\kpc$)          & ($\kpc$)   \\
\noalign{\smallskip}
\hline
\noalign{\smallskip}
\multispan{3}{\bf{Fiducial Model}:} & \multispan7{$n_{\rm p}=15$~~$n_{\rm dof}=46$~~$\chi^2_{\rm min}=55.1$~~$\langle\sigma_i\rangle=0.5\arcsec$~~$\frac{\rm Pr(model|data,PIEMD)}{\rm Pr(ABDE|data,PIEMD)}=1$\hfil}\\
\noalign{\smallskip}
&Halo A  & ~~~ &  $0.4^{+1.5}_{-1.1}$ & $2.5^{+1.1}_{-1.5}$  & $0.44^{+0.1}_{-0.07}$    & $123.5\pm1.5$   & $1243^{+60}_{-62}$    & $137^{+11}_{-14}$ & $1000$  \\
\noalign{\smallskip}
&Halo B  & ~~~ &  $+27.8$            & $-32.2$             & $0.0$                   & ...                   & $428^{+41}_{-46}$     & $50$   & $1000$  \\
\noalign{\smallskip}
&Halo D  & ~~~ &  $-20.8$            & $+48.1$             & $0.0$                   & ...                   & $441^{+114}_{-100}$      & $50$  & $1000$  \\
\noalign{\smallskip}
&Halo E  & ~~~ &  $+18.7$            & $+101.3$            & $0.23$                  & $23.8$                & $454^{+153}_{-85}$      & $35^{+39}_{-15}$  & $1000$  \\
\noalign{\smallskip}
&BCG     & ~~~ &  $0.0$              & $0.0$               & $0.20$                  & $124$                 & $231^{+38}_{-26}$        & $<2$    & $78^{+23}_{-36}$ \\
\noalign{\smallskip}
&$L^\star$ galaxy & ~~~ & ... & ... & ... & ... & $180$ & $0.2$ & $30$ \\
\noalign{\smallskip}
&A4 & ~~~ & \multispan3{$z=2.5\pm0.2$\hfil}\\
\noalign{\smallskip}
&A6 & ~~~ & \multispan3{$z=1.7\pm0.5$\hfil}\\
\noalign{\smallskip}
&A7 & ~~~ & \multispan3{$z\ge1.4$\hfil}\\
\noalign{\smallskip}
\noalign{\smallskip}
\multispan{3}{\bf{ABCDE Model:}} & \multispan7{$n_{\rm p}=16$~~$n_{\rm dof}=45$~~$\chi^2_{\rm min}=~~55.1$~~$\langle\sigma_i\rangle=0.5\arcsec$~~$\frac{\rm Pr(model|data,PIEMD)}{\rm Pr(ABDE|data,PIEMD)}=4\times10^{-2}$\hfil}  \\
\noalign{\smallskip}
\multispan{3}{\bf{ABD Model:}} & \multispan7{$n_{\rm p}=13$~~$n_{\rm dof}=48$~~$\chi^2_{\rm min}=471.3$~~$\langle\sigma_i\rangle=1.4\arcsec$~~$\frac{\rm Pr(model|data,PIEMD)}{\rm Pr(ABDE|data,PIEMD)}=2\times10^{-96}$\hfil} \\
\noalign{\smallskip}
\multispan{3}{\bf{ABE Model:}} & \multispan7{$n_{\rm p}=14$~~$n_{\rm dof}=47$~~$\chi^2_{\rm min}=~~86.8$~~$\langle\sigma_i\rangle=0.6\arcsec$~~$\frac{\rm Pr(model|data,PIEMD)}{\rm Pr(ABDE|data,PIEMD)}=4\times10^{-12}$\hfil}\\
\noalign{\smallskip}
\multispan{3}{\bf{ADE Model:}} & \multispan7{$n_{\rm p}=14$~~$n_{\rm dof}=47$~~$\chi^2_{\rm min}=110.9$~~$\langle\sigma_i\rangle=0.8\arcsec$~~$\frac{\rm Pr(model|data,PIEMD)}{\rm Pr(ABDE|data,PIEMD)}=8\times10^{-13}$\hfil}\\
\noalign{\smallskip}
\multispan{3}{\bf{AB Model:}} & \multispan7{$n_{\rm p}=12$~~$n_{\rm dof}=49$~~$\chi^2_{\rm min}=493.8$~~$\langle\sigma_i\rangle=1.5\arcsec$~~$\frac{\rm Pr(model|data,PIEMD)}{\rm Pr(ABDE|data,PIEMD)}=6\times10^{-93}$\hfil}\\
\noalign{\smallskip}
\multispan{3}{\bf{AD Model:}} & \multispan7{$n_{\rm p}=12$~~$n_{\rm dof}=49$~~$\chi^2_{\rm min}=539.5$~~$\langle\sigma_i\rangle=1.7\arcsec$~~$\frac{\rm Pr(model|data,PIEMD)}{\rm Pr(ABDE|data,PIEMD)}=2\times10^{-102}$\hfil}\\
\noalign{\smallskip}
\multispan{3}{\bf{AE Model:}} & \multispan7{$n_{\rm p}=13$~~$n_{\rm dof}=48$~~$\chi^2_{\rm min}=130.7$~~$\langle\sigma_i\rangle=0.9\arcsec$~~$\frac{\rm Pr(model|data,PIEMD)}{\rm Pr(ABDE|data,PIEMD)}=4\times10^{-20}$\hfil}\\
\noalign{\smallskip}
\multispan{3}{\bf{A Model:}} & \multispan7{$n_{\rm p}=11$~~$n_{\rm dof}=50$~~$\chi^2_{\rm min}=551.1$~~$\langle\sigma_i\rangle=1.8\arcsec$~~$\frac{\rm Pr(model|data,PIEMD)}{\rm Pr(ABDE|data,PIEMD)}=5\times10^{-107}$\hfil}\\
\noalign{\smallskip}
\hline
\label{tab:models}
\end{tabular}
\end{center}
\end{table*}

\subsection{Mass and Structure of the Cluster Core}\label{sec:mass}

We use the fiducial model to compute the projected mass within a
projected radius of $R=500\kpc$ and the fraction of that mass residing
in the groups and the cluster galaxy population, obtaining a mass of
$M(\le500\kpc)=(6.7\pm0.4)\times10^{14}\Msol$ and a substructure
fraction of $\fsub(\le500\kpc)=0.25\pm0.12$.  MACS\,J1149 therefore
has a mass and substructure fraction comparable with the most
disturbed of the clusters studied at $z\simeq0.2$ by \citet{Smith05a}
and \citet{Richard09}, see also \citet{Smith08}.  However despite
these global similarities, we note that none of the lower redshift
clusters contained three group-scale halos.  MACS\,J1149 is therefore
the most complex strong-lensing cluster studied to date.

Despite having similar velocity dispersions, the three group-scale
halos have different optical morphologies, suggesting that they may
have suffered different infall histories.  Luminous structure B
appears as an extended finger pointing SE from the cluster center
(Fig.~\ref{fig:arcs}) dominated by two bright elliptical galaxies: one
on which we placed halo B in the lens model, and a second one
$\sim20''$ further SE.  Luminous structure D comprises a dense group
of galaxies, none of which dominate the optical luminosity.  Finally,
luminous structure E is dominated by a single bright elliptical
galaxy.  Despite our ability to find an acceptable fit, we therefore
expect that improvements to the strong-lensing constraints, especially
spectroscopic redshifts of A6 and A7 will help to constrain in more
detail the structure of these ``groups'', aided by more flexible
modeling schemes \citep[e.g.][]{Jullo09}.

\begin{figure}
\centerline{\psfig{file=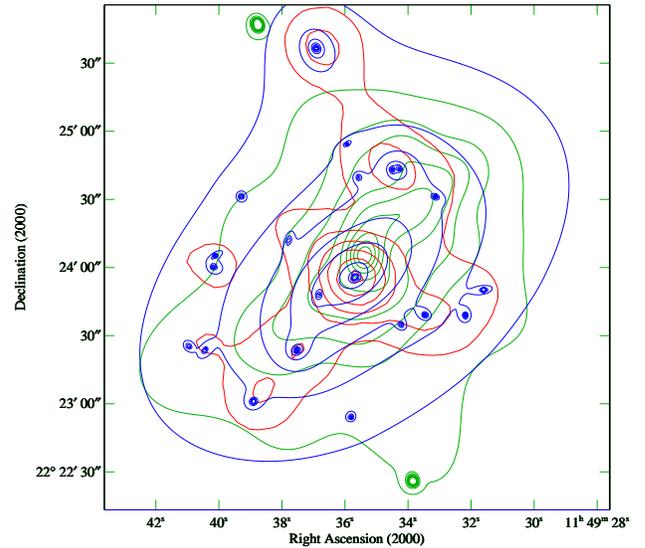,width=80mm,angle=0}}
\caption{ The central $\sim1.5\Mpc\times1.5\Mpc$ of MACS\,J1149, as
  revealed by the adaptively smoothed luminosity density of cluster
  galaxies (red), projected total mass map calculated from the
  gravitational lens model (blue) and adaptively smoothed X-ray
  surface brightness contours from \emph{Chandra} observations
  (green).  All contours are spaced linearly. 
\label{fig:massmap}
}
\end{figure}

In Fig.~\ref{fig:massmap} we show the contours of luminosity density
(\S\ref{sec:lum}), isomass density (\S\ref{sec:model};
Table~\ref{tab:models}), and X-ray surface brightness from
\emph{Chandra} observations \citep{Ebeling07}.  By construction the
mass contours agree well with the luminosity density contours, except
that luminous structure C is not embedded in an extended dark matter
halo.  The peak of the X-ray emission is offset from the BCG by
$\sim15\arcsec$, and the overall X-ray morphology is elongated in a
NW--SE direction, i.e.\ in the same direction as the mass and light
contours.  MACS\,J1149 therefore follows the well-established trend
for X-ray luminous clusters with multi-modal mass distributions to
have an X-ray morphology that is not centered on the BCG
\citep[][]{Smith05a,Poole06,Powell09,Sanderson09}, adding weight to
the conclusion that this is a merging cluster.

\subsection{ Intrinsic Properties of the Spiral Galaxy at z=1.491 }

We use the fiducial lens model (Table~\ref{tab:models}) to calculate
the intrinsic properties of the multiply-imaged disk-galaxy at
$z=1.491$.  A1.1, A1.2, A1.3, and A1.4 are magnified gravitationally
by $\mu=23$, $18$, $8$, and $23$ respectively, with a typical
uncertainty of $30\%$.  The unlensed apparent magnitude of this galaxy
in the $I$-band is therefore $I\simeq23.4\pm0.3$, corresponding to
$M_B\simeq-20.7$ in the rest frame.  A1 is therefore comparable with
$L^\star$ galaxies in the local universe \citep{Norberg02}, the
faintest of the lensed disk galaxies in \citeauthor{Swinbank06}'s
(\citeyear{Swinbank06}) study of the Tully Fisher relation at
$z\simeq1$, and the lensed Sa galaxy at $z=1.6$ studied by
\cite{Smith02b}.  In contrast, it is $\sim1\mag$ fainter than
previously studied unlensed disk galaxies at similar redshifts
\citep{vanDokkum01,Wright09}.  Many of these galaxies are bulge
dominated systems; in contrast, A1 is not dominated by its bulge, with
a rest frame B-band bulge-to-total ratio of $B/T\sim0.4-0.5$.

We also estimate the star formation rate from the observed
$V_{555}$-band flux -- adopting the calibration of \cite{Kennicutt98},
we obtain a global star formation rate of $\sim6\Msolpyr$.  Individual
HII regions in local galaxies typically span $\sim50-100\pc$
\citep{Gonzalez97}, which translates to $\sim6-12\mas$ at $z=1.491$.
The lens magnification of $\mu=23$ suffered by A1.1 boosts the angular
scale of HII regions in this galaxy to $\sim30-60\mas$, bringing them
within reach of instruments such as OSIRIS on Keck and NIFS on Gemini
North.  This galaxy therefore offers a unique opportunity to study the
distribution of star formation, chemical abundance gradients, and even
the physics of individual star-forming regions at a look back time of
$9.3\Gyr$ at a level of detail similar to that achieved at $z=0.1$.

\subsection{ Density Profile }\label{sec:zitrin}

\begin{figure}
\centerline{\psfig{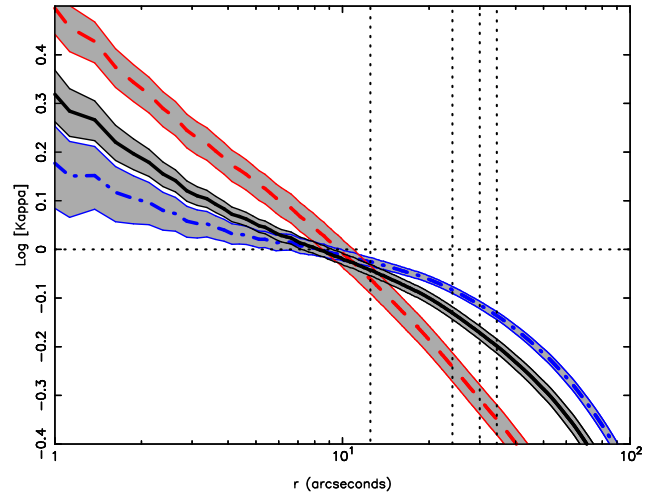}}
\caption{ The density profile of MACS\,J1149 from
  our fiducial model (black solid), our ZB-constrained model with all
  redshifts as free parameters (red dashed), and the latter with the
  redshifts of A1, A2, A3, A4 fixed (blue, dot-dashed), as 
  described in \S\ref{sec:zitrin}.  In each case the grey filled
  regions show the $95\%$ confidence interval around the
  best-fit model.  The horizontal line marks the critical density
  required for strong lensing ($\kappa=1$), and the vertical
  dotted-dashed lines mark the average cluster-centric radius at which
  (from left to right) images of A1, A2, A3, and A4 are observed.  
\label{fig:prof}
}
\end{figure}

We show the density profile from our fiducial model in
Fig.~\ref{fig:prof}.  Parameterizing the profile as $\kappa\sim
r^\gamma$, we obtain $\gamma\simeq-0.3\pm0.05$ in the radial range
$r\sim3-30\arcsec$.  During peer review of this letter,
\citet[][hereafter ZB]{Zitrin09b} claimed that the density profile of
MACS\,J1149 is flat and critically convergent ($\kappa\simeq1$) out to
$r\sim200\kpc$, equivalent to $\sim30\arcsec$.  Inspection of ZB's
Fig.~5 reveals that the average slope of their profile at
$r\sim3-30\arcsec$ is $\gamma\simeq-0.1\pm0.02$ ($68\%$ confidence,
assuming their error bars are $1\sigma$).  We therefore rule out ZB's
model at $\sim7\sigma$ confidence; a flat model (i.e.\ $\gamma=0$)
within this radial range is ruled out at $12\sigma$.

The two main differences between ZB's model and ours are the
following.  First, ZB's multiple-image interpretation is different
from ours: they claim to find a fifth image of A1, plus six additional
multiple-image systems (their 5--10), all of which do not pass the
strict criteria described in \S\ref{sec:model}; in addition, ZB do not
identify our systems A6 and A7.  Second, ZB's analysis contains no
spectroscopic or photometric redshift information for their multiple
images and, more importantly, ZB do not treat these unknown redshifts
as free parameters in their model -- their model contains just 6 free
parameters, which describe the cluster mass distribution.

We attempt to reproduce ZB's flat profile by fitting a model to all of
their multiple image identifications, including the putative fifth
image of system A1 (marked by a black circle in the A1.2 panel of
Fig. 1).  We treat the redshifts of all multiple images as free
parameters. The resulting best-fit ``ZB-constrained'' model has an
image-plane rms of $\langle\sigma_i\rangle=1.2\arcsec$ , i.e.\ more
than twice that of our fiducial model, dominated by the fifth image of
A1 and ZB's systems 5--10.  The density profile associated with this
model is shown in Fig.~\ref{fig:prof} and is in fact {\em steeper}
than ours.  However, once redshifts are no longer included as free
parameters in the fit, but fixed at values that differ, to varying
degrees, from the true, measured values, the sensitivity of the
density profile to a chosen set of fixed redshift values becomes
apparent.  We demonstrate this by setting the redshifts of A1, A2, A3,
and A4 to $z=1.5$, $1.6$, $1.8$, and $1.8$, i.e.\ to values that are
permitted by the model uncertainties but are in fact not the measured
ones. The density profile (Fig.~\ref{fig:prof}) resulting from these
erroneous assumptions is nearly flat, with $\gamma\sim-0.14$.

We conclude that ZB's claim of a flat density profile is highly
sensitive to the details of the method by which they chose to assign
fixed redshifts to multiple-image systems.  These problems may have
been compounded by mis-identification of some multiple image systems.

\section{ Summary }\label{sec:summary}

We have presented new \emph{HST}/ACS and Keck I/LRIS observations of
MACS\,J1149, a massive X-ray selected galaxy cluster at $z=0.544$
discovered in the Massive Cluster Survey.  These data reveal seven
robustly identified multiply-imaged galaxies, three of which we have
confirmed spectroscopically.  The most spectacular system is a
multiply-imaged face-on disk galaxy at $z=1.491$ that we identify as
an $L^\star$ ($M_B\simeq-20.7$) late-type ($B/T\ls0.5$) galaxy with an
ongoing star formation rate of $\sim6\Msolpyr$; the brightest images
of this galaxy are magnified by $\mu=23$.  Future observations using
integral field spectrographs should probe its properties in exquisite
detail, thanks to the combination of lens magnification and fortuitous
viewing angle.

We use the positions and redshifts of robustly identified
multiply-imaged galaxies to constrain a detailed model of the mass and
structure of the cluster core.  Our fiducial model contains the main
cluster halo plus three group-scale halos; the probability of a model
this complex, relative to less complex models is ${\rm P}(N_{\rm
  halo}=4)/{\rm P}(N_{\rm halo}<4)\ge10^{12}$ where $N_{\rm halo}$ is
the number of cluster/group-scale halos.  We measure the mass and
fraction of mass residing in substructures to be
$M(\le500\kpc)=6.7\pm0.4\times10^{14}\Msol$ and
$\fsub(\le500\kpc)=0.25\pm0.12$ respectively.  In summary, MACS\,J1149
is the most complex strong-lensing cluster core studied to date, its
relatively dis-assembled nature being qualitatively consistent with
the expectation that clusters at high redshifts are on average less
mature than those at lower redshifts.  A more complete view will
emerge from our analysis of the full sample of MACS clusters at
$z>0.5$ (Smith et al., in prep.).

We also obtain a power law density profile slope of
$\gamma=-0.3\pm0.05$ ($95\%$ confidence error bars) on scales of
$r\sim3-30\arcsec$, thereby ruling out density profile slopes as flat
as those recently proposed by \cite{Zitrin09b} at $\gs7\sigma$
confidence.  In summary, \citeauthor{Zitrin09b}'s result can be
explained by an absence of multiple-image redshifts of any form in
their study, and by them not treating the unknown redshifts as free
parameters in their model.  These issues are probably compounded by
them mis-identifying some multiple-image systems.  Overall, this
underlines the critical importance of measuring spectroscopic
redshifts of multiply-imaged galaxies for reliable lens models of
strong lensing clusters.

\section*{ Acknowledgments }

GPS thanks Keren Sharon and Phil Marshall for assistance with the Keck
observations, and acknowledges Paul May, and Chris Berry for helpful
discussions.  GPS and IRS acknowledge support from the Royal Society
and STFC.  AMS acknowledges a RAS Fellowship.  HE, JPK and GPS
acknowledge support from STScI under grant GO-09722.  ML acknowledges
the Centre National d'Etudes Spatiales (CNES) and the Danish National
Research Foundation for their support.  JPK acknowledges support from
CNRS.


\begin{thebibliography}{36}
\expandafter\ifx\csname natexlab\endcsname\relax\def\natexlab#1{#1}\fi

\bibitem[{{Borys} {et~al.}(2004){Borys}, {Chapman}, {Donahue}, {Fahlman},
  {Halpern}, {Kneib}, {Newbury}, {Scott}, \& {Smith}}]{Borys04}
{Borys}, C., {Chapman}, S., {Donahue}, M., {Fahlman}, G., {Halpern}, M.,
  {Kneib}, J.-P., {Newbury}, P., {Scott}, D., \& {Smith}, G.~P. 2004, \mnras,
  352, 759

\bibitem[{{Bradley} {et~al.}(2008){Bradley}, {Bouwens}, {Ford}, {Illingworth},
  {Jee}, {Ben{\'{\i}}tez}, {Broadhurst}, {Franx}, {Frye}, {Infante}, {Motta},
  {Rosati}, {White}, \& {Zheng}}]{Bradley08}
{Bradley}, L.~D., {Bouwens}, R.~J., {Ford}, H.~C., {Illingworth}, G.~D., {Jee},
  M.~J., {Ben{\'{\i}}tez}, N., {Broadhurst}, T.~J., {Franx}, M., {Frye}, B.~L.,
  {Infante}, L., {Motta}, V., {Rosati}, P., {White}, R.~L., \& {Zheng}, W.
  2008, \apj, 678, 647

\bibitem[{{Broadhurst} {et~al.}(2005){Broadhurst}, {Ben{\'{\i}}tez}, {Coe},
  {Sharon}, {Zekser}, {White}, {Ford}, {Bouwens}, {Blakeslee}, {Clampin},
  {Cross}, {Franx}, {Frye}, {Hartig}, {Illingworth}, {Infante}, {Menanteau},
  {Meurer}, {Postman}, {Ardila}, {Bartko}, {Brown}, {Burrows}, {Cheng},
  {Feldman}, {Golimowski}, {Goto}, {Gronwall}, {Herranz}, {Holden}, {Homeier},
  {Krist}, {Lesser}, {Martel}, {Miley}, {Rosati}, {Sirianni}, {Sparks},
  {Steindling}, {Tran}, {Tsvetanov}, \& {Zheng}}]{Broadhurst05acs}
{Broadhurst}, T., {Ben{\'{\i}}tez}, N., {Coe}, D., {Sharon}, K., {Zekser}, K.,
  {White}, R., {Ford}, H., {Bouwens}, R., {Blakeslee}, J., {Clampin}, M.,
  {Cross}, N., {Franx}, M., {Frye}, B., {Hartig}, G., {Illingworth}, G.,
  {Infante}, L., {Menanteau}, F., {Meurer}, G., {Postman}, M., {Ardila}, D.~R.,
  {Bartko}, F., {Brown}, R.~A., {Burrows}, C.~J., {Cheng}, E.~S., {Feldman},
  P.~D., {Golimowski}, D.~A., {Goto}, T., {Gronwall}, C., {Herranz}, D.,
  {Holden}, B., {Homeier}, N., {Krist}, J.~E., {Lesser}, M.~P., {Martel},
  A.~R., {Miley}, G.~K., {Rosati}, P., {Sirianni}, M., {Sparks}, W.~B.,
  {Steindling}, S., {Tran}, H.~D., {Tsvetanov}, Z.~I., \& {Zheng}, W. 2005,
  \apj, 621, 53

\bibitem[{{Comerford} {et~al.}(2006){Comerford}, {Meneghetti}, {Bartelmann}, \&
  {Schirmer}}]{Comerford06}
{Comerford}, J.~M., {Meneghetti}, M., {Bartelmann}, M., \& {Schirmer}, M. 2006,
  \apj, 642, 39

\bibitem[{{Ebeling} {et~al.}(2007){Ebeling}, {Barrett}, {Donovan}, {Ma},
  {Edge}, \& {van Speybroeck}}]{Ebeling07}
{Ebeling}, H., {Barrett}, E., {Donovan}, D., {Ma}, C.-J., {Edge}, A.~C., \&
  {van Speybroeck}, L. 2007, \apjl, 661, L33

\bibitem[{{Ebeling} {et~al.}(2001){Ebeling}, {Edge}, \& {Henry}}]{Ebeling01}
{Ebeling}, H., {Edge}, A.~C., \& {Henry}, J.~P. 2001, \apj, 553, 668

\bibitem[{{Ebeling} {et~al.}(2006){Ebeling}, {White}, \&
  {Rangarajan}}]{Ebeling06}
{Ebeling}, H., {White}, D.~A., \& {Rangarajan}, F.~V.~N. 2006, \mnras, 368, 65

\bibitem[{{Ellis} {et~al.}(2001){Ellis}, {Santos}, {Kneib}, \&
  {Kuijken}}]{Ellis01}
{Ellis}, R., {Santos}, M.~R., {Kneib}, J., \& {Kuijken}, K. 2001, \apjl, 560,
  L119

\bibitem[{{Franx} {et~al.}(1997){Franx}, {Illingworth}, {Kelson}, {van Dokkum},
  \& {Tran}}]{Franx97}
{Franx}, M., {Illingworth}, G.~D., {Kelson}, D.~D., {van Dokkum}, P.~G., \&
  {Tran}, K. 1997, \apjl, 486, L75

\bibitem[{{Gladders} {et~al.}(2002){Gladders}, {Yee}, \&
  {Ellingson}}]{Gladders02}
{Gladders}, M.~D., {Yee}, H.~K.~C., \& {Ellingson}, E. 2002, \aj, 123, 1

\bibitem[{{Gonzalez Delgado} \& {Perez}(1997)}]{Gonzalez97}
{Gonzalez Delgado}, R.~M. \& {Perez}, E. 1997, \apjs, 108, 199

\bibitem[{{Inada} {et~al.}(2003){Inada}, {Oguri}, {Pindor}, {Hennawi}, {Chiu},
  {Zheng}, {Ichikawa}, {Gregg}, {Becker}, {Suto}, {Strauss}, {Turner},
  {Keeton}, {Annis}, {Castander}, {Eisenstein}, {Frieman}, {Fukugita}, {Gunn},
  {Johnston}, {Kent}, {Nichol}, {Richards}, {Rix}, {Sheldon}, {Bahcall},
  {Brinkmann}, {Ivezi{\'c}}, {Lamb}, {McKay}, {Schneider}, \& {York}}]{Inada03}
{Inada}, N., {Oguri}, M., {Pindor}, B., {Hennawi}, J.~F., {Chiu}, K., {Zheng},
  W., {Ichikawa}, S.-I., {Gregg}, M.~D., {Becker}, R.~H., {Suto}, Y.,
  {Strauss}, M.~A., {Turner}, E.~L., {Keeton}, C.~R., {Annis}, J., {Castander},
  F.~J., {Eisenstein}, D.~J., {Frieman}, J.~A., {Fukugita}, M., {Gunn}, J.~E.,
  {Johnston}, D.~E., {Kent}, S.~M., {Nichol}, R.~C., {Richards}, G.~T., {Rix},
  H.-W., {Sheldon}, E.~S., {Bahcall}, N.~A., {Brinkmann}, J., {Ivezi{\'c}}, {\v
  Z}., {Lamb}, D.~Q., {McKay}, T.~A., {Schneider}, D.~P., \& {York}, D.~G.
  2003, \nat, 426, 810

\bibitem[{{Jullo} \& {Kneib}(2009)}]{Jullo09}
{Jullo}, E. \& {Kneib}, J.-P. 2009, \mnras, 395, 1319

\bibitem[{{Jullo} {et~al.}(2007){Jullo}, {Kneib}, {Limousin},
  {El{\'{\i}}asd{\'o}ttir}, {Marshall}, \& {Verdugo}}]{Jullo07}
{Jullo}, E., {Kneib}, J.-P., {Limousin}, M., {El{\'{\i}}asd{\'o}ttir}, {\'A}.,
  {Marshall}, P.~J., \& {Verdugo}, T. 2007, New Journal of Physics, 9, 447

\bibitem[{{Kennicutt}(1998)}]{Kennicutt98}
{Kennicutt}, R.~C. 1998, \araa, 36, 189

\bibitem[{{Kneib} {et~al.}(2004){Kneib}, {Ellis}, {Santos}, \&
  {Richard}}]{Kneib04b}
{Kneib}, J., {Ellis}, R.~S., {Santos}, M.~R., \& {Richard}, J. 2004, \apj, 607,
  697

\bibitem[{{Kneib} {et~al.}(1996){Kneib}, {Ellis}, {Smail}, {Couch}, \&
  {Sharples}}]{Kneib96}
{Kneib}, J.-P., {Ellis}, R.~S., {Smail}, I., {Couch}, W.~J., \& {Sharples},
  R.~M. 1996, \apj, 471, 643

\bibitem[{{Limousin} {et~al.}(2007){Limousin}, {Richard}, {Jullo}, {Kneib},
  {Fort}, {Soucail}, {El{\'{\i}}asd{\'o}ttir}, {Natarajan}, {Ellis}, {Smail},
  {Czoske}, {Smith}, {Hudelot}, {Bardeau}, {Ebeling}, {Egami}, \&
  {Knudsen}}]{Limousin07a1689}
{Limousin}, M., {Richard}, J., {Jullo}, E., {Kneib}, J.-P., {Fort}, B.,
  {Soucail}, G., {El{\'{\i}}asd{\'o}ttir}, {\'A}., {Natarajan}, P., {Ellis},
  R.~S., {Smail}, I., {Czoske}, O., {Smith}, G.~P., {Hudelot}, P., {Bardeau},
  S., {Ebeling}, H., {Egami}, E., \& {Knudsen}, K.~K. 2007, \apj, 668, 643

\bibitem[{{Ma} {et~al.}(2008){Ma}, {Ebeling}, {Donovan}, \& {Barrett}}]{Ma08}
{Ma}, C.-J., {Ebeling}, H., {Donovan}, D., \& {Barrett}, E. 2008, \apj, 684,
  160

\bibitem[{{Norberg} {et~al.}(2002){Norberg}, {Cole}, {Baugh}, {Frenk},
  {Baldry}, {Bland-Hawthorn}, {Bridges}, {Cannon}, {Colless}, {Collins},
  {Couch}, {Cross}, {Dalton}, {De Propris}, {Driver}, {Efstathiou}, {Ellis},
  {Glazebrook}, {Jackson}, {Lahav}, {Lewis}, {Lumsden}, {Maddox}, {Madgwick},
  {Peacock}, {Peterson}, {Sutherland}, \& {Taylor}}]{Norberg02}
{Norberg}, P., {Cole}, S., {Baugh}, C.~M., {Frenk}, C.~S., {Baldry}, I.,
  {Bland-Hawthorn}, J., {Bridges}, T., {Cannon}, R., {Colless}, M., {Collins},
  C., {Couch}, W., {Cross}, N.~J.~G., {Dalton}, G., {De Propris}, R., {Driver},
  S.~P., {Efstathiou}, G., {Ellis}, R.~S., {Glazebrook}, K., {Jackson}, C.,
  {Lahav}, O., {Lewis}, I., {Lumsden}, S., {Maddox}, S., {Madgwick}, D.,
  {Peacock}, J.~A., {Peterson}, B.~A., {Sutherland}, W., \& {Taylor}, K. 2002,
  \mnras, 336, 907

\bibitem[{{Ofek} {et~al.}(2008){Ofek}, {Seitz}, \& {Klein}}]{Ofek08}
{Ofek}, E.~O., {Seitz}, S., \& {Klein}, F. 2008, \mnras, 389, 311

\bibitem[{{Oke} {et~al.}(1995){Oke}, {Cohen}, {Carr}, {Cromer}, {Dingizian},
  {Harris}, {Labrecque}, {Lucinio}, {Schaal}, {Epps}, \& {Miller}}]{oke95}
{Oke}, J.~B., {Cohen}, J.~G., {Carr}, M., {Cromer}, J., {Dingizian}, A.,
  {Harris}, F.~H., {Labrecque}, S., {Lucinio}, R., {Schaal}, W., {Epps}, H., \&
  {Miller}, J. 1995, \pasp, 107, 375

\bibitem[{{Poole} {et~al.}(2006){Poole}, {Fardal}, {Babul}, {McCarthy},
  {Quinn}, \& {Wadsley}}]{Poole06}
{Poole}, G.~B., {Fardal}, M.~A., {Babul}, A., {McCarthy}, I.~G., {Quinn}, T.,
  \& {Wadsley}, J. 2006, \mnras, 373, 881

\bibitem[{{Powell} {et~al.}(2009){Powell}, {Kay}, \& {Babul}}]{Powell09}
{Powell}, L.~C., {Kay}, S.~T., \& {Babul}, A. 2009, ArXiv e-prints

\bibitem[{{Richard} {et~al.}(2009){Richard}, {Smith}, et al}]{Richard09}
{Richard}, J., {Smith}, G.~P., {Kneib}, J.-P., {Ellis}, R.~S., et al
2009, MNRAS, submitted

\bibitem[{{Sand} {et~al.}(2008){Sand}, {Treu}, {Ellis}, {Smith}, \&
  {Kneib}}]{Sand08}
{Sand}, D.~J., {Treu}, T., {Ellis}, R.~S., {Smith}, G.~P., \& {Kneib}, J.-P.
  2008, \apj, 674, 711

\bibitem[{{Sanderson} {et~al.}(2009){Sanderson}, {Edge}, \&
  {Smith}}]{Sanderson09}
{Sanderson}, A.~J.~R., {Edge}, A.~C., \& {Smith}, G.~P. 2009, \mnras, 398, 1698

\bibitem[{{Sharon} {et~al.}(2005){Sharon}, {Ofek}, {Smith}, {Broadhurst},
  {Maoz}, {Kochanek}, {Oguri}, {Suto}, {Inada}, \& {Falco}}]{Sharon05}
{Sharon}, K., {Ofek}, E.~O., {Smith}, G.~P., {Broadhurst}, T., {Maoz}, D.,
  {Kochanek}, C.~S., {Oguri}, M., {Suto}, Y., {Inada}, N., \& {Falco}, E.~E.
  2005, \apjl, 629, L73

\bibitem[{{Smail} {et~al.}(2007){Smail}, {Swinbank}, {Richard}, {Ebeling},
  {Kneib}, {Edge}, {Stark}, {Ellis}, {Dye}, {Smith}, \& {Mullis}}]{Smail07}
{Smail}, I., {Swinbank}, A.~M., {Richard}, J., {Ebeling}, H., {Kneib}, J.-P.,
  {Edge}, A.~C., {Stark}, D., {Ellis}, R.~S., {Dye}, S., {Smith}, G.~P., \&
  {Mullis}, C. 2007, \apjl, 654, L33

\bibitem[{{Smith} {et~al.}(2005){Smith}, {Kneib}, {Smail}, {Mazzotta},
  {Ebeling}, \& {Czoske}}]{Smith05a}
{Smith}, G.~P., {Kneib}, J.-P., {Smail}, I., {Mazzotta}, P., {Ebeling}, H., \&
  {Czoske}, O. 2005, \mnras, 359, 417

\bibitem[{{Smith} {et~al.}(2002){Smith}, {Smail}, {Kneib}, {Davis}, {Takamiya},
  {Ebeling}, \& {Czoske}}]{Smith02b}
{Smith}, G.~P., {Smail}, I., {Kneib}, J.-P., {Davis}, C.~J., {Takamiya}, M.,
  {Ebeling}, H., \& {Czoske}, O. 2002, \mnras, 333, L16

\bibitem[{{Smith} \& {Taylor}(2008)}]{Smith08}
{Smith}, G.~P. \& {Taylor}, J.~E. 2008, \apjl, 682, L73

\bibitem[{{Swinbank} {et~al.}(2006){Swinbank}, {Bower}, {Smith}, {Smail},
  {Kneib}, {Ellis}, {Stark}, \& {Bunker}}]{Swinbank06}
{Swinbank}, A.~M., {Bower}, R.~G., {Smith}, G.~P., {Smail}, I., {Kneib}, J.-P.,
  {Ellis}, R.~S., {Stark}, D.~P., \& {Bunker}, A.~J. 2006, \mnras, 368, 1631

\bibitem[{{Swinbank} {et~al.}(2007){Swinbank}, {Bower}, {Smith}, {Wilman},
  {Smail}, {Ellis}, {Morris}, \& {Kneib}}]{Swinbank07}
{Swinbank}, A.~M., {Bower}, R.~G., {Smith}, G.~P., {Wilman}, R.~J., {Smail},
  I., {Ellis}, R.~S., {Morris}, S.~L., \& {Kneib}, J.-P. 2007, \mnras, 376, 479

\bibitem[{{van Dokkum} \& {Stanford}(2001)}]{vanDokkum01}
{van Dokkum}, P.~G. \& {Stanford}, S.~A. 2001, \apjl, 562, L35

\bibitem[{{Wright} {et~al.}(2008){Wright}, {Larkin}, {Law}, {Steidel},
  {Shapley}, \& {Erb}}]{Wright09}
{Wright}, S.~A., {Larkin}, J.~E., {Law}, D.~R., {Steidel}, C.~C., {Shapley},
  A.~E., \& {Erb}, D.~K. 2008, ArXiv e-prints

\bibitem[{{Zitrin} \& {Broadhurst}(2009)}]{Zitrin09b}
{Zitrin}, A. \& {Broadhurst}, T. 2009, \apjl, 703, L132

\end{thebibliography}

\end{document}